\documentstyle[12pt]{article}

\begin{document}

{\bf SYMMETRY CONSTRAINTS AND THE ELECTRONIC STRUCTURES OF A QUANTUM DOT
WITH THIRTEEN ELECTRONS}\vspace{1pt}

\ \ \ \ \ \

G.M. Huang, Y.M. Liu, and C.G. Bao\vspace{1pt}

\ \ \ \ \ \ \ \ \ \ \ \

The State Key Laboratory of Optoelectronic Materials and Technologies, and

Department of Physics, Zhongshan University, \ Guangzhou, 510275, P.R. China%
\vspace{1pt}

\vspace{1pt}

ABSTRACT: The symmetry constraints imposing on the quantum states of a dot
with 13 electrons has been investigated. Based on this study, the favorable
structures (FSs) of each state has been identified. Numerical calculations
have been performed to inspect the role played by the FSs. It was found
that, if a first-state has a remarkably competitive FS, this FS would be
pursued and the state would be crystal-like and have a specific core-ring
structure associated with the FS. The magic numbers are found to be closely
related to the FSs.\vspace{1pt}

PACS(numbers): 73.61.-r\vspace{1pt}

\vspace{1pt}

1, INTRODUCTION \qquad

\vspace{1pt}\qquad Modern experimental techniques, e.g., by using
electrostatic gates and by etching, allow a certain number of electrons to
be confined in semiconductor heterostructures.$^{1-6}$\ \ Such many-electron
systems have much in common with atoms, yet they are man-made structures and
are usually called '' quantum dot ''. \ The number of electrons contained in
a dot ranges from a few to a few thousands, they are confined in a domain
one hundred or more times larger than the atoms. Thus, in addition to atoms,
nuclei,$\cdot \cdot \cdot $ that exist in nature, quantum dots as a new kind
of system will definitely contain new and rich physics, and therefore they
attract certainly the interest of academic research.

\qquad\ \ On the other hand, the properties of the dots can be changed \ in
a controlled way, e.g., by changing the gate voltage or by applying an
adjustable magnetic field, etc. Therefore, these systems \ have a great
potential in application. Due to this fact, the investigation of quantum
dots is a hot topic in recent years$^{1-6}$. \ In the experimental aspect,
progress has been made to reveal different kinds of physical property. A
crucial point is to clarify the electronic structures. An important step
along this line is the first observation of the Coulomb blockade spectra via
the measurement \ of conductance as a function of gate voltage$^{7}$, where
very clear level structure has been demonstrated. Afterwards, a substantial
amount of information is drawn from conductance measurement. \ The
measurement of the difference in chemical potential exhibits also clear
shell structures$^{8}$. \ The excitation of electron can be probed by
far-infrared and capacitance spectroscopy.$^{9,10}$ \ With further progress
in experimental techniques, the dots will definitely be understood better
and better, and they will serve as a rich source of information on many-body
physics in the coming years.

\qquad In the theoretical aspects, detailed information on electronic
structures has been obtained for the systems with a smaller N (say, N$<$10)$%
^{2,4,6}$. \ When N is small, the effect of symmetry was found to be very
important, e.g., the magic angular momenta of few-electron dots originate
from the constraint of symmetry$^{11-13}$. When N is larger (say, N $\geq $%
10), the effect of symmetry is scarcely studied. The systems with a larger N
are themselves very attractive, because they might possess both the features
of few-body and many-body systems. Thus the understanding of these systems
might serve as a bridge to connect few-body and many-body physics. In a
previous paper, the electronic structures of a dot with nine electrons have
been studied$^{13}$. The present paper is a continuation of the previous
one, and is dedicated to the study of the dot with N=13 and with the spins
polarized. The choice of thirteen is rather arbitrary, just because it is
explicitly larger than the systems with N$<$10 which have already been
extensively studied, and because it is not very large so that accurate
numerical calculations (in the qualitative sense) and detailed analysis can
still be performed. From a previous study by a number of authors$^{13-18}$ ,
it is believed that a general picture of dots would consist of a core
surrounding by a ring. It would be interesting to see, \ when N is larger,
how the details of the core-ring structure would be and how these structures
are affected by symmetry . Such a study would exhibit further insight of
many-body physics.

\qquad In what follows, the 13-body Schr\"{o}dinger equation is solved \ via
an exact diagonalization of the Hamiltonian, the accuracy has been
evaluated. The underlying dynamical and symmetry background has been
studied. \ Favorable structures for each state have been suggested based on
symmetry consideration. The eigenwavefunctions have been analyzed in detail
to exhibit how the electronic structures are affected by symmetry. The
appearance of magic numbers is discussed.\vspace{1pt}

\ \ \ \

2, HAMILTONIAN AND THE APPROACH

\vspace{1pt}\qquad Let the electrons be fully polarized (therefore the
spin-part can be neglected and the spatial wave functions are totally
antisymmetric) , and confined in a 2-dimensional plane by a parabolic
confinement. The Hamiltonian reads

\ \ \ \ \ \ \ \ \ \ \ \ \ \ \ $H=T+U\qquad \qquad (1.1)$

$\qquad T=-\sum\limits_{j=1}^{N}\frac{\hbar ^{2}}{2m^{\ast }}\nabla
_{j}^{2}\qquad \qquad \qquad (1.2)$

$\qquad U=\sum\limits_{j=1}^{N}\frac{1}{2}m^{\ast }\omega _{o}^{2}r_{j}^{2}+%
\frac{e^{2}}{4\pi \varepsilon _{r}\varepsilon _{0}}\sum\limits_{j<k}^{N}%
\frac{1}{r_{jk}}$ \ \ \ \ \ \ \ \ \ \ \ \ \ \ \ \ \ \ \ \ \ \ \ \ \ \ \ \ \
(1.3)

where m* is the effective electron mass, $\varepsilon _{r}$ is the
dielectric constant, and $\hbar \omega _{o}$measures the strength of the
parabolic confinement\ (\ $\omega _{o}$\ arises mainly from a magnetic field
$B$. This field leads also to a term linearly proportional to $B$ . This
term has been neglected because it does not at all affect the
eigenwavefunctions, and therefore not affect the electronic structures).

\qquad\ In order to diagonalize the Hamiltonian, a set of orthonormalized
single-particle harmonic oscillation (h.o.)states $\phi _{mk}$ \ are
introduced. \ Here, \ $\phi _{mk}$ \ is an eigenstate of a pure h.o.
Hamiltonian

\qquad\ \ \ \ $h=-\frac{\hbar ^{2}}{2m^{\ast }}\nabla ^{2}+\frac{1}{2}%
m^{\ast }\Omega _{0}^{2}r^{2}$ \ \ \ \ \ \ \ \ \ \ \ \ \ \ \ \ \ \ \ \ \ \ \
\ \ \ \ \ \ \ \ \ \ \ \ \ \ \ \ \ \ \ \ \ \ \ \ \ \ \ \ \ \ \ \ \ \ \ \ \ \
\ \ \ \ \ \ \ \ \ \ \ \ (2)

where $\Omega _{0}$ is an adjustable parameter in general not equal to $%
\omega _{o}$, This eigenstate has eigenenergy $(m+k+1)\hbar \Omega _{0}$ \ \
and angular momentum $(m-k)\hbar $ \ \ . From them the many-body basis
functions (BFs)

\bigskip\ $\ \psi _{\alpha }(1,2,\cdot \cdot \cdot ,N)=\frac{1}{\sqrt{N!}}%
\left|
\begin{array}{cccc}
\phi _{m_{1}k_{1}}(\vec{r}_{1}) & \phi _{m_{1}k_{1}}(\vec{r}_{2}) & \cdot
\cdot \cdot & \phi _{m_{1}k_{1}}(\vec{r}_{N}) \\
\phi _{m_{2}k_{2}}(\vec{r}_{1}) & \phi _{m_{2}k_{2}}(\vec{r}_{2}) & \cdot
\cdot \cdot & \phi _{m_{2}k_{2}}(\vec{r}_{N}) \\
\cdot \cdot \cdot & \cdot \cdot \cdot & \cdot \cdot \cdot & \cdot \cdot \cdot
\\
\phi _{m_{N}k_{N}}(\vec{r}_{1}) & \phi _{m_{2}k_{2}}(\vec{r}_{2}) & \cdot
\cdot \cdot & \phi _{m_{N}k_{N}}(\vec{r}_{N})
\end{array}
\right| $\ \ \ \ \ \ \ \ \ (3)

with a given total orbital angular momentum $L=\sum_{i}(m_{i}-k_{i})$ are
composed.

\ \ \ \ From the BFs, the eigenstates of the dot are expanded as

\ \ \ \ \ \ \ \ \ \ \ \ \ \ \ \ \ \ \ \ \ \ \ \ \ \ $\Psi =\sum C_{\alpha
}\psi _{\alpha }$ \ \ \ \ \ \ \ \ \ \ \ \ \ \ \ \ \ \ \ \ \ \ \ \ \ \ \ \ \
\ \ \ \ \ \ \ \ \ \ \ \ \ \ \ \ \ \ \ \ \ \ \ \ \ \ \ \ \ \ \ \ \ \ \ \ \ \
\ \ \ (4)

\ \ \ \ \ where \ the coefficients $C_{\alpha }$ can be obtained via a
procedure of diagonalization. \ The $\ \psi _{\alpha }$ are arranged in such
a sequence that $<\psi _{\alpha }|H|\psi _{\alpha }>$\ $\leq $ $<\psi
_{\alpha +1}|H|$\ $\psi _{\alpha +1}>$\ . Evidently, in such a sequence, the
$\ \psi _{\alpha }$ with a smaller $\alpha $ is more important to the
low-lying states, while those with a very large $\alpha $ can be neglected.
\ The $H$ will be diagonalized step by step. In the first step, $H$ is
diagonalized in a smaller space with N$_{a}$ BFs ( $\psi _{1}$ to $\psi
_{N_{a}}$) . Then, $H$ is diagonalized\ again in a larger space with N$_{b}$
BFs ( from $\psi _{1}$ to $\psi _{N_{b}}$ , and N$_{b}$ is considerably
larger than N$_{a}$) . This process repeats again and again until a
satisfactory convergency of the lower eigenenergies is achieved. \ In the
first step, all the $\ \psi _{\alpha }$ for the diagonalization is limited
to the lowest Landau levels (LLL), \ i.e., all the $\phi _{m_{i}k_{i}}$
contained in $\ \psi _{\alpha }$\ have $k_{i}=0$. However, step by step, BFs
of higher Landau levels will mixed in. In order to speed up the convergency,
the $\Omega _{0}$\ in eq.(2) is considered as a variational parameter to
optimize the lower eigenenergies emerged from the diagonalization.

\vspace{1pt}\label{w} \ \ \ \ \ \ \ \ \ In the following calculation, we
have m*=0.067$m_{e}$, $\hbar \omega _{0}$=3meV, $\varepsilon _{r}$=12.4 (for
a GaAs dot). \ To show the convergency, as an example, the lowest
eigenenergies with $L$=82 are obtained as $436.895,$ $436.806$ and $436.760$%
meV when the number of BFs are 6000, 9000, and 12000, respectively. One can
see that the convergency is not very good. However, the densities calculated
below by using 6000, 9000 and 12000 BFs are indistinguishable (e.g. in
Fig.1). Since we are mainly interested in the qualitative aspect, the
accuracy that we have achieved is sufficient.

\vspace{1pt} \ \ \ \ \ \ \ \ \ After the diagonalization the eigenstates are
obtained. The series of states \ having the same $L$ is labeled as ($L$)$%
_{i} $. The $i=1$ state (the lowest of the $L$-series) is called a
first-state.

\qquad The eigenwavefunctions of a 13-electron system are complicated. In
order to extract informations from them the following physical quantities
are defined and calculated. They are the one-body density

$\rho _{1}(r_{1})=\int \left| \Psi _{L}\right| ^{2}d{\bf r}_{2}d{\bf r}%
_{3}\cdot \cdot \cdot d{\bf r}_{13}$ \ \ \ ,\ \ \ \ \ \ \ \ \ \ \ \ \ \ \ \
\ \ \ \ \ \ \ (5a)

the two-body density

$\rho _{2}({\bf r}_{1}{\bf ,r}_{2})=\int \left| \Psi _{L}\right| ^{2}d{\bf r}%
_{3}d{\bf r}_{4}\cdot \cdot \cdot d{\bf r}_{13}$ \ \ \ ,\ \ \ \ \ \ \ \ \ \
\ \ \ \ \ \ \ \ \ \ (5b)

\vspace{1pt}and the three-body density

$\rho _{3}({\bf r}_{1}{\bf ,r}_{2}{\bf ,r}_{3})=\int \left| \Psi _{L}\right|
^{2}d{\bf r}_{4}d{\bf r}_{5}\cdot \cdot \cdot d{\bf r}_{13}$ \ ,\ \ \ \ \ \
\ \ \ \ \ \ \ \ \ \ (5c)

\qquad It was found that in many cases the $\rho _{1}(r)$ has an outer peak
and an inner peak with a minimum lying in between (at $r=a)$. In this case
we can define an outer region ($r$ $\geq a$) and an inner region ($r$ $<$ $a$%
) . \ Accordingly, we can define the average number of particles $N_{out}$
and $N_{in}$ contained in the outer and inner regions, respectively, \ as

\qquad $N_{out}=N\int_{a}^{\infty }\rho _{1}(r_{1})d{\bf r}_{1}$ \ \ \ \ \ \
\ \ \ \ \ \ \ \ \ \ \ \ \ \ (6a)\ \ \

$\qquad N_{in}=N\int_{0}^{a}\rho _{1}(r_{1})d{\bf r}_{1}$ \ \ \ \ \ \ \ \ \
\ \ \ \ \ \ \ \ \ \ \ \ (6b)

For example, \ the (88)$_{1}$ state has $a$ =367.8$\stackrel{o}{A}$, $%
N_{out}=$ 9.97 and $N_{in}=$3.03.

\qquad Once the border $a$ is defined, we can define the angular momenta $%
l_{out}$ and the moments of inertia $I_{out}$\ contributed by the outer
region, respectively ,\ as

\qquad $l_{out}=N\int_{a}^{\infty }d{\bf r}_{1}\int \Psi _{L}^{\ast }%
\widehat{l_{1}}\Psi _{L}d{\bf r}_{2}\cdot \cdot \cdot d{\bf r}_{13}$ \ \ \ \
\ \ \ \ (7)\ \ \

and

\ \ \ $\ \ I_{out}$ $=M\int_{a}^{\infty }\rho _{1}(r_{1})r_{1}^{2}d{\bf r}%
_{1}$\ \ \ \ \ \ \ \ \ \ \ \ \ \ \ \ \ \ \ \ \ (8)\

\ where $M=Nm^{\ast }$\ is the total mass. Similarly, the $l_{in}$ and $%
I_{in}$ contributed from the inner region can also be defined. \ Although
these quantities are not good quantum numbers, they can help us to
understand better the physics as shown later.

\vspace{1pt}

3, DYNAMICAL AND\ SYMMETRY BACKGROUND

\qquad Quantum mechanic systems are affected by both dynamical reasons and
symmetry consideration . The following points are noticeable:

\ \ \ \ \ \ (i) {\bf Core-ring structures}.

The spatial wave functions of low-lying states are mainly distributed in an
area where the total potential energy $U$ (eq.(1.3)) is lower. \ In
particular, they would like to be distributed surrounding the (local)minima
of $U$. In order to find out the (local)minima, let \ N$_{in}$ electrons be
contained inside to form a core, and N$_{out}$ electrons be contained
outside to form a ring, N$_{in}$+N$_{out}=$N. When the relative locations of
the electrons are appropriately adjusted (e.g., they form two homocentric
regular polygons with or without an electron at the center) $U$ will be
optimized and arrives at its (local) minimum $U_{opt}$, \ the associated
configuration is called an N$_{in}-$N$_{out}$ core-ring configuration. In
this configuration, let the \ ratio of the radii of the outer\ polygon and
the inner polygon be denoted as G$_{opt}$. $U_{opt}$ and \ G$_{opt}$ are
given in Table 1.\vspace{1pt}

\ \ Table 1, \ The optimal values $U_{opt}$ and the associated G$_{opt}$ of
the (local)minima of $U$ , each is associated with a N$_{in}-$N$_{out}$
core-ring configuration.

\begin{tabular}[t]{|c|c|c|c|c|c|c|c|c|}
\hline
N$_{in}-$N$_{out}$ & 1-12 & 2-11 & \ 3-10 & 4-9 & 5-8 & 6-7 & 7-6 & 8-5 \\
\hline
$U_{opt}$ (meV) & 281.41 & 278.35 & 274.83 & 274.22 & 274.93 & 276.31 &
275.73 & 278.76 \\ \hline
G$_{opt}$ &  & 3.73 & 2.83 & 2.41 & 2.16 & 2.02 & 1.77 & 1.70 \\ \hline
\end{tabular}
\ \ \

\ \ \ \ \ \ \ \ \ \

\vspace{1pt}Evidently, a too small or too large N$_{out}$ (say, N$_{out}\leq
7$ or N$_{out}\geq 11)$\ is not advantageous to binding. Furthermore, the
outer polygon should be neither too close to nor too far away from the core.

\qquad In what follows, when the wave function of a state is distributed
surrounding a N$_{in}-$N$_{out}$ core-ring configuration, then the state is
said to have a N$_{in}-$N$_{out}$ structure. If the configuration has an
electron at the center, then the structure is further denoted as (N$%
_{in})_{c}-$N$_{out}$ , \ otherwise as\ (N$_{in})_{h}-$N$_{out}$ . The
subscript $h$ implies a hollow structure.

\qquad It is shown in the table that the $U_{opt}$ of a number of
configurations are quite close to each other. At a first glance, one might
expect that a strong mixing of geometric configurations would occur and
would spoil the crystal-like picture. However, this is not true mainly due
to the quantum constraints as we shall see later.

\vspace{1pt}

\qquad (ii) {\bf Uniform rotation} .

\qquad\ Let us consider first a classical model system of two rotating
homocentric rings. the outer ring has ($b\leq r\leq a$), while the inner
ring has ($d\leq r\leq c$ , and $\ $c$\leq b$). \ Let the angular momentum,
the moment of inertia and the angular velocity of the outer (inner) ring be $%
l_{out}$, $I_{out}$ and $\omega _{out}$ ( $l_{in}$, $I_{in}$\ and \ $\omega
_{in}$), respectively. The total angular momentum $L=l_{out}+l_{in}=I_{out}%
\omega _{out}+I_{in}\omega _{in}$ , and the total rotation energy $T=\frac{1%
}{2}(I_{out}\omega _{out}^{2}+I_{in}\omega _{in}^{2})$ . Now, let us ask how
the $\omega _{out}$ and $\omega _{in}$ would be chosen so that $T$ is
minimized under the condition that $L$ is conserved? The answer is simply $%
\omega _{out}=\omega _{in}=L/(I_{out}+I_{in})=L/I$. This fact implies that
if the two rings are rotating with the same angular velocity, the rotation
energy can be reduced. Although this point is viewed from classical
mechanics, however the first-states of a quantum mechanic system would do
its best to lower the energy, thus they would pursue a uniform rotation,
i.e., $\omega _{out}\approx \omega _{in}$ .

\qquad From the point of view of quantum mechanics, the low-lying states are
mainly dominated by the BFs belonging to the LLL. In these BFs, all the
single-particle state $\phi _{mk}$ have \ $k=0$ and angular momentum $%
l=m-k=m $. For each $\phi _{mk}$ , the angular velocity can be defined as $%
\omega =<l>/(m^{\ast }<r^{2}>)$, which is proportional to $\frac{l}{l+1}$ if
$k=0$. \ Evidently, $\omega $ is close to a constant unless $l$ is very
small. Thus , for the BFs of the LLL, all the electrons rotate with similar
angular velocities, and we have the uniform rotation $\omega _{out}\approx
\omega _{in}$\ .

\qquad (iii) {\bf Symmetry constraints and the \ favorable structures}.

It has been found that{\it \ inherent nodal surfaces are imposed in wave
functions by symmetry, thereby the structures of quantum states are
seriously affected}.$^{12,19-21}$ In the case of 2-dimensional polarized
quantum dots , it was found that a wave function would be zero when the
electrons locate at the vertexes of a regular\ N-side polygon if $L\neq $N$%
(j+\frac{1+(-1)^{N}}{4})$ , where $j$ is an integer$^{6,11,22}$. This
constraint can be generalized to the core-ring structures. Let the ring has
an angular momentum $l_{out}$ , while the core has $l_{in}$ . When the outer
particles locate at the vertexes of a N$_{out}$-side polygon, and the inner
particles locate at the vertexes of a N$_{o}$-side polygon (N$_{o}$=N$_{in}$
or N$_{in}-1$, in the latter case an electron would stay at the center) ,
then it is straight forward to prove that the wave function would be zero if

$l_{out}\neq $N$_{out}(j_{2}+(1+(-1)^{{N}_{out}})/4)$ $\qquad (9a)$

or

$l_{in}\neq $N$_{o}(j_{1}+(1+(-1)^{{N}_{o}}/4)$ $\qquad (9b)$

where $j_{1}$ and $j_{2}$ are integers. In other words, the above
configuration would be prohibited if $l_{out}($ $l_{in})$ does not relate to
N$_{out}$ (N$_{o}$) in the above way. \ {\it Thus, a (N}$_{in})-${\it N}$%
_{out}${\it \ structure would be pursued by a first-state only if the }$L$%
{\it \ can be divided as a sum of\ }$l_{in}${\it \ and }$l_{out}${\it \ so
that \ the requirements (9a) and (9b) are fulfilled. }\ If this happens, the
(N$_{in})-$N$_{out}$ structure is called a candidate of favorable structure%
{\it \ (CFS)} of the state. \ Incidentally, for an eigenstate, both the $%
l_{out}$ and $l_{in}$ are not good quantum numbers, they appear as the
angular momenta of the main component of eigenwavefunctions.

\qquad \qquad\ Usually each state may have a number of CFS, some of them are
not competitive due to having a too small or too large N$_{out}$, they can
be neglected. In what follows, among the CFS of a state, if some of them
have N$_{out}\geq 8$, then those with N$_{out}\leq 7$ are neglected; if all
the CFS has N$_{out}\leq 7,$ then all of them would be neglected except the
one with the largest N$_{out}$; all the CFS with N$_{out}=12$ are neglected
without exception. After the neglect, the remaining CFS \ are call the
favorable structures (FSs), they are listed in Table 2. E.g., the $L=86$
state has four CFS , \ namely the (5)$_{c}$-8, (9)$_{c}$-4, (11)$_{c}$-2,
and (12)$_{c}$-1 . \ Among them only the (5)$_{c}$-8 as a FS is listed in
Table 2. \ \ When a state has more than three FS, only the most competitive
three are listed.

\vspace{1pt}

\qquad (iv) {\bf \ Excitation of the core}

This paper concerns only the low-lying states with $L\geq N(N-1)$ $/2$ ( or
the filling factor $\nu \leq 1$ ), they contain mainly the BFs belonging to
the LLL. In these BFs, the angular momenta of any pair of electrons can not
be the same due to the Pauli Principle. Therefore, they can be denoted as $%
\psi _{\alpha }=\{l_{1}l_{2}\cdot \cdot \cdot l_{N}\}$ with \ $l_{i}<l_{i+1}$%
.

\qquad\ Since we have

\vspace{1pt}\qquad $<\phi _{l0}|r^{2}|\phi _{l0}>=(l+1)\frac{\hbar }{m^{\ast
}\Omega _{o}},\qquad \qquad (10)$

the spatial distribution of the wave function $\phi _{l0}$ depends on $l$. \
The smaller the $l$, the smaller the size. \ Thus, for the $\psi _{\alpha }$
belonging to the LLL, $l_{in}$\ is \ just equal to $l_{1}+l_{2}+\cdot \cdot
\cdot +l_{{N}_{in}}$ . \ If $l_{1}=0$, \ there must be an electron staying
at the center, because the $\phi _{00}$ wave function is distributed closely
surrounding the center. \ Thus, a (N$_{in})_{c}$-N$_{out} $ structure must
be contributed by the $\psi _{\alpha }$ with $l_{1}=0$\ , \ while a (N$%
_{in})_{h}$-N$_{out}$ structure is contributed by those with $l_{1}>0$\ . \
When all the $l_{i}$ of the inner electrons satisfies \ $l_{i}+1=l_{i+1}$\ ,
the inner electrons are said to be compactly aligned. Meanwhile, $l_{in}$
would arrive at its lower bound $(l_{in})_{b}=$N$_{in}($N$_{in}-1)$ $/2$ ,
if $l_{1}=0$. In this case, we say that the core is inert (not excited).
Otherwise, we have $l_{in}>$ $(l_{in})_{b}$, and we say that the core is
excited. Evidently, all the hollow states must have $l_{1}>0$, thus they
have an excited core.

\qquad When $L$ $\leq 90$, core excitation is not possible (unless the
electrons jump to higher LLL), therefore the first-states would have a (N$%
_{in})_{c}-$N$_{out}$ structure with the core inert. However, when $L>90$,
core excitation might occur. It implies two cases: (a) The inner electrons
have their $l_{i}$ remaining to be compactly aligned but with $l_{1}=k$ ,
and therefore have a (N$_{in})_{h}$-N$_{out}$ hollow structure . \ (b) The $%
l_{i}$ \ of the core are no longer aligned compactly, e.g., $l_{1}=0$ while $%
l_{2}=2$, etc. .

\qquad It was found that, when $L$ is not large (say, $L\leq 101$ ), the
first-states have either an inert core or an excited compact core with $%
l_{1}=1$ , as shown in Table 2 . However, when $L$ is large, higher core
excitation with $l_{1}>1$ will emerge as shown later.

\qquad Incidentally, \ due to eq.(10), the compact alignment of the angular
momenta also implies a compact alignment of radial positions. Thus, in the
core-ring structures, the groups of inner and outer electrons may each
compactly aligned. The associated \ BF can be simply denoted as $%
\{l_{1}-l_{N_{in}},l_{N_{in}+1}-l_{N}\}$ (e.g., \{1,2,3, 6,7,$\cdot \cdot
\cdot $,15\} $\equiv $\{1-3,6-15\}.). This is called a two-bunched BF by Ruan%
$^{23}$ (The one-bunched BF \{$l_{1}-l_{N}$\} is a special case of
two-bunched BF with $l_{N_{in}}=l_{N_{in}+1}-1$ ). It is straight forward to
prove that, for a CFS of a $\nu \leq 1$\ state, \ among the BFs of the CFS,
one and only one of them is a two-bunched BF belonging to the LLL. \ Thus, a
simple way to find out the CFS is to look for the two-bunched BFs of a state.

\qquad

(v) {\bf Particle separation}

\qquad It is noted that the $U$ in eq.(1.3) can be exactly rewritten as

\qquad $U=\frac{1}{2}M\omega _{o}^{2}R_{c}^{2}+\sum\limits_{j<k}
u_{eff}(r_{jk})\qquad \qquad \qquad (11.1)$

Where $R_{c}$ is the radial distance of the c.m., and $u_{eff}(r_{jk})$ is
the effective pairwise interaction

\qquad $u_{eff}(r_{jk})=\frac{(m\ast \omega _{o})^{2}}{2M}r_{jk}^{2}+\frac{%
e^{2}}{4\pi \varepsilon _{r}\varepsilon _{0}}\cdot \frac{1}{r_{jk}}\qquad
\qquad \qquad (11.2)$

There is a minimum in $u_{eff}$ located at $r_{jk}=r_{u}=(\frac{e^{2}M}{4\pi
\varepsilon _{r}\varepsilon _{0}(m\ast \omega _{o})^{2}})^{1/3}$\ .
Evidently, if each electron separates from all its adjacent electrons by
this distance, the potential energy can be minimized. \ Therefore, in
low-lying states, \ adjacent electrons would roughly keep the separation $%
r_{u}$ . \ With the above parameters, $r_{u}=576.3\stackrel{\circ }{A}$\ \
.The $r_{u}$ is a basic measure and is useful for the understanding of
electronic correlation and the size of the system. \ Obviously, for a (N$%
_{in})-$N$_{out}$ structure, the distance between the ring and the core
depends closely on $r_{u}$ . \

\qquad

(vi) {\bf Core-ring separation}

\qquad It is recalled that, in order to minimize the potential energy, the
ring should separate from the core by an appropriate distance. In this
subsection, we shall evaluate the core-ring separation by using the
approximation of uniform rotation.

\qquad\ For a given CFS with the given N$_{out}$ and $l_{out}$ , let us
define a quantity

\qquad $\stackrel{\_}{g}=\sqrt{\frac{l_{out}/l_{in}}{N_{out}/N_{in}}}\qquad
\qquad (12)$

On the other hand, we have

$I_{out}=m^{\ast }N_{out}<r^{2}>_{ring}\qquad \qquad (13)$

(this equation is the same as eq.(8)), and a similar equation for $I_{in}$.
Thus we have

$\stackrel{\_}{g}=\sqrt{\frac{l_{out}/l_{in}}{I_{out}/I_{in}}\cdot
(<r^{2}>_{ring}/<r^{2}>_{core})}=\sqrt{\frac{\omega _{out}}{\omega _{in}}%
\cdot (<r^{2}>_{ring}/<r^{2}>_{core})}\qquad \qquad (14)$\vspace{1pt}

It is believed that the uniform rotation is a good approximation for the
first-states, because they should do their best to lower the energy (this is
a point remain to be checked). Under this approximation

\qquad $\stackrel{\_}{g}\approx (<r^{2}>_{ring}/<r^{2}>_{core})^{1/2}\qquad
\qquad (15)$

The optimal value of the right hand side of eq.(15) has been denoted as G$%
_{opt}$ given in Table 1. Thus, if a FS has its $\stackrel{\_}{g}$
(evaluated from the definition eq.(12)) close to G$_{opt}$ , then the
core-ring separation is appropriate and the FS is advantageous to binding
and therefore competitive. Otherwise, it is not.

\ The $\stackrel{\_}{g}/$G$_{opt}$ of the FS are also listed in Table 2,
many of them are found to be very close to one. E.g., the FS of the (86)$%
_{1} $ is a (5)$_{c}-$8 structure with $\stackrel{\_}{g}$ =2.18, \ the
associated G$_{opt}$ is 2.16 (cf. Table 1) , thus they are close to each
other .

\qquad The above points are important to the following discussion. \vspace{%
1pt}\ {\it When a first-state has a FS which is superior than the other FSs
(or the state has only one FS), the FS is expected to be dominant. In this
case the state would have a clear geometric feature arising from the N}$%
_{in}-${\it N}$_{out}${\it \ structure of the FS, and appear to be
crystal-like.} However, when a first-state has a few nearly equally
competitive FSs, its structure can not be uniquely predicted. Nevertheless,
the Table 2 is a key to understand the electronic structures.\vspace{1pt}

\ \ \ \ \ \ \ \ Table 2, \ Characters of the first-state from symmetry
consideration and from our calculations.

\begin{tabular}{|l|l|l|l|l|l|l|l|l|l|}
\hline
& \multicolumn{5}{|l|}{Favorable structures and} & \multicolumn{4}{|l|}{
Quantities evaluated from} \\
& \multicolumn{5}{|l}{their \ related \ features} & \multicolumn{4}{|l|}{the
$\rho _{1}$ of the first-states} \\ \hline
L & FS & $l_{1}$ & $l_{in}$ & $\overline{g}$ & $\overline{g}/$G$_{opt}$ & $a$
& $N_{in}$ & $l_{in}$ & $\gamma $ \\ \hline
81 & (10)$_{c}$-3 & 0 & 45 & 1.6330 &  &  &  &  &  \\ \hline
82 & (9)$_{c}$-4 & 0 & 36 & 1.6956 &  &  &  &  &  \\ \hline
83 & (8)$_{c}$-5 & 0 & 28 & 1.7728 & 1.04 &  &  &  &  \\ \hline
84 & (7)$_{c}$-6 & 0 & 21 & 1.8708 & 1.05 & 2.700 & 6.55 & 21.30 & 0.89 \\
\hline
85 & (6)$_{c}$-7 & 0 & 15 & 2.0000 & 0.99 & 2.550 & 5.76 & 16.45 & 0.91 \\
\hline
86 & (5)$_{c}$-8 & 0 & 10 & 2.1794 & 1.01 & 2.288 & 4.65 & 10.52 & 0.93 \\
\hline
87 & (4)$_{c}$-9 & 0 & 6 & 2.4495 & 1.02 & 2.100 & 3.85 & 7.21 & 0.95 \\
\hline
88 & (3)$_{c}$-10 & 0 & 3 & 2.9155 & 1.03 & 1.889 & 3.03 & 4.44 & 0.96 \\
\hline
89 & (2)$_{c}$-11 & 0 & 1 & 4.0000 & 1.07 & 1.555 & 2.00 & 1.90 & 0.98 \\
\hline
90 & (1)$_{c}$-12 & 0 & 0 &  &  & 1.120 & 0.97 & 0.42 & 1.094 \\ \hline
92 & (6)$_{c}$-7 & 0 & 15 & 2.0976 & 1.04 & 2.700 & 6.26 & 19.70 & 0.94 \\
\hline
93 & (8)$_{c}$-5 & 0 & 28 & 1.9272 & 1.13 & 2.414 & 4.99 & 12.43 & 0.964 \\
\hline
94 & (5)$_{c}$-8 & 0 & 10 & 2.2913 & 1.06 & 2.377 & 5.01 & 12.51 & 0.967 \\
\hline
95 & (9)$_{h}$-4 & 0 & 45 & 1.5811 &  & 2.205 & 4.06 & 8.30 & 0.967 \\ \hline
96 & (4)$_{c}$-9 & 0 & 6 & 2.5820 & 1.07 & 2.181 & 3.89 & 7.38 & 0.998 \\
\hline
97 & (7)$_{h}$-6 & 1 & 28 & 1.6956 & 0.96 & 1.942 & 3.04 & 4.80 & 0.967 \\
\hline
98 & (3)$_{c}$-10 & 0 & 3 & 3.0822 & 1.08 & 1.926 & 2.91 & 4.19 & 1.012 \\
\hline
99 & (5)$_{h}$-8 & 1 & 15 & 1.8708 & 0.86 & 1.519 & 1.74 & 1.86 & 0.904 \\
\hline
100 & (2)$_{c}$-11 & 0 & 1 & 4.2426 & 1.14 & 1.611 & 1.90 & 1.98 & 0.981 \\
\hline
100 & (4)$_{h}$-9 & 1 & 10 & 2.0000 & 0.83 &  &  &  &  \\ \hline
101 & (3)$_{h}$-10 & 1 & 6 & 2.1794 & 0.77 & 1.936 & 2.82 & 6.54 & 0.813 \\
\hline
\end{tabular}

4, EIGENENERGIES

\vspace{1pt}\qquad After performing the diagonalization, eigenenergies and
eigenstates are obtained. Let $E((L)_{i})$ be the energy of the $(L)_{i}$ \
state. It is noted that , for a first-state, if the Coulomb repulsion among
the electrons are removed, all the electrons would fall in the LLL with the
energy $(L+N)\hbar \omega _{o}$ . \ For this reason, let us define $%
\varepsilon (L)\equiv E((L)_{1})-(L+N)\hbar \omega _{o}.$This quantity is a
measure of the Coulomb repulsion in the first-states, which is plotted in
Fig.2 in accord with $L$. When $L$ increases, the size of the system will
increase a little , \ the Coulomb repulsion will thereby decrease. Thus, $%
\varepsilon (L)$ decreases monotonously with $L$ as shown in the figure. \
However, there are four platforms. We shall return to this point later.

\vspace{1pt}

5, ELECTRONIC STRUCTURES $\left( 78\leq L\leq 90\right) $

\qquad In what follows mainly the results of the first-states are given. We
use $a_{M}\equiv \sqrt{\frac{\hbar }{m^{\ast }\omega _{o}}}=$ 194.71$%
\stackrel{\circ }{A}$ \ as the unit of length. The optimal separation $%
r_{u}= $ 2.96$a_{M}$ .

\qquad Let us begin from the state with the filling factor $\nu =1$ , namely
the (78)$_{1}$ state. This state has only one BF \{0,1,2,$\cdot \cdot \cdot $%
12\} (for short, \{0-12\}) belonging to the LLL, this BF has a weight\
85.5\%. In this BF, the electrons are roughly uniformly distributed inside a
circle as shown in Fig.3a. \ It is noted that a clear geometric structure
arises from the coherent mixing of BFs. Due to the lack of mixing, the (78)$%
_{1}$ can not have a clear geometric structure, therefore it is liquid-like
as shown in Fig.4a.

\qquad\ On the other hand, for the number N together with two arbitrary
integers $n$ ($\leq $N), and $j^{\prime }$, there is an identity

\vspace{1pt}\qquad $\frac{N(N-1)}{2}+j^{\prime }N=\frac{n(n+2j^{\prime }-1)}{%
2}+\frac{(N-n)(N+n+2j^{\prime }-1)}{2}\qquad \qquad \qquad (16)$

Let the left hand side be equal to $L$, and the two terms at the right be
equal to $l_{in}$ \ and $l_{out}$. Then this identity is associated with a
division of $L$. When $j^{\prime }=0$ , the left hand side of (16) is equal
to 78 . It is easy to see that the pair N$_{o}=n-1$ and $l_{in}$ meet the
requirement of eq.(9b), while the pair N$_{out}$=$N-n$ and $l_{out}$ meet
the requirement of eq.(9a). \ Thus, eq.(16) implies that all the (N$%
_{o}+1)_{c}$-N$_{out}$ structures with N$_{o}=0$ to 12 are the CFS of the $%
L=78$ states. Therefore the wave function of the (78)$_{1}$can get access to
all the \ symmetric configurations$^{12,13}$, and thus is nodeless (except a
pair of electrons overlap with each other). Accordingly, the energy of this
state is lower.

\qquad For the (79)$_{1}$ state, there is also only one BF \{0-11,13\}
belonging to the LLL. Thus this state is also liquid-like as shown in
Fig.4b. \ However, on the contrary with the (78)$_{1}$ , all the (N$_{in})$-N%
$_{out}$ structures are not the CFS of the (79)$_{1}$, except the (12)$_{c}$%
-1 which is very poor in binding. Thus the energy of this state is much
higher. Owing to the (78)$_{1}$ \ is lower while the (79)$_{1}$ is higher,
the difference leads to a platform appearing in Fig.2 between $L=78$ and 79.

\qquad Ranging from \ (79)$_{1}$ to (90)$_{1}$, all these states have only
one FS, thus their structures can be well predicted. \ The N$_{out}$ of
their FS (cf. Table 2) increases from 1 to 12, this leads to a regular
variation of their electronic structure. When N$_{out}$ is small (say, N$%
_{out}\leq 5$ ), the outward electrons are found to be very close to the
core. As a result, their ring-core-structures are ambiguous as shown in
Fig.3b and 4b, where the patterns are representative for the (79)$_{1}$ to
(83)$_{1}$ states.\ \ In these states the FS itself is not competitive. \
This fact would lead to a stronger mixing of structures, and therefore they
are liquid-like.

\qquad\ Even in the liquid-like states, electronic correlation can still be
viewed via the three-body densities as shown in Fig.5a and 5b, they are
representative. Fig.5a for the (81)$_{1}$ exhibits that the three outward
electrons (two are labelled by white spots and one by a double-peak, which
implies an oscillation around an equilibrium position) are very close \ to
the core. This fact supports the presumption that the FS, namely the (10)$%
_{c}$-3 structure (cf. Table 2), is pursued by the state . Although the $U$
of the (10)$_{c}$-3 is higher, however no other better symmetric
configurations are allowed by symmetry. Consequently, the component of the
(10)$_{c}$-3 is still relatively important . \ Since the outward electrons
are so close, the core is strongly deformed. \ There are three peaks at the
outer ridge of the core, it implies that three inward electrons form a
regular triangle close to the border. \ Fig.5b for the (82)$_{1}$ exhibits
that the four outward electrons are also very close \ to the core. This fact
supports again that the FS is pursued. \ The core is also strongly deformed
with four inward electrons forming a square close to the border. \

\qquad\ The pursuit of the FS can also be viewed by observing the
composition of the wave functions. For the (81)$_{1}$ , the BF with the
largest weight (35.4\%) is the \{0-9,11-13\}, in which the electrons are
divided into two compact bunchs, and therefore supports directly the (10)$%
_{c}$-3 structure . For the (82)$_{1}$\ , the BF with the largest weight
(33.4\%) is the \{0-8,10-13\}. \

\qquad When $L\geq 84$, the N$_{out}$ of the FS is $\geq 6$. Since the
outward electrons would separate (roughly by $r_{u}$) from each other, a
larger N$_{out}$ definitely leads to a \ larger ring. Consequently, the
outward electrons are no more close to the core, and the ring-core structure
becomes explicit. This is shown in Fig.3c to 3f for the (84)$_{1}$ to (90)$%
_{1}$ states, where the outward peak becomes larger and larger. \ The point $%
\ a$ separating the inner and outer regions can be well defined.
Accordingly, \ the quantities related to eq.(6) to (8) can be calculated as
listed \ in Table 2. In particular, a quantity related to the uniformity of
rotation

\qquad $\gamma =\frac{l_{out}}{I_{out}}/\frac{l_{in}}{I_{in}}=\omega
_{out}/\omega _{in}\qquad \qquad (17)$

is defined and is also listed.

\vspace{1pt}\qquad It is exhibited in Table 2 that, in the range \ 84$\leq
L\leq 90$, $a$ and $N_{in}$ are decreasing . This coincides with the
reduction of the core of the FS. In particular, the N$_{in}$ of the FS are
one-to-one close to the $N_{in}$ from calculation. This fact confirms that
the FSs are pursued by the first-states. In general the $N_{in}$ and $l_{in}$
deviate more or less from those of the FS, this is due to the mixing of the
FS together with other minor structures ( the inner electrons may
occasionally go out ,or the core may get slightly excited). \ E.g., the wave
function of the (87)$_{1}$\ has $N_{in}=3.85$ and $l_{in}=7.21$ , while its
FS has N$_{in}$=4 and $l_{in}$=$(l_{in})_{b}=$6 (incidentally, a
core-excitation may cause a big increase of $l_{in}$ ). \ Furthermore, the $%
\gamma $ are close to the unity, it implies that the rotation is roughly
uniform. However, the slight deviation of $\gamma $ implies that the system
is not entirely rigid.

\qquad It is recalled that the $\rho _{2}$ of the $L\leq 83$ first-states
appear as liquid-like. However, when N$_{out}$ is neither very small nor
very large (say, 6 $\leq $N$_{out}\leq 10$ ), the $U$ of the core-ring
structure is lower, and thereby the associated FS becomes more dominant.
This would lead to a clear crystal-like picture as shown in Fig.4c to 4g,
where the outward electrons form a regular polygon. The number of vertexes
(from 6 to 10) is just equal to the N$_{out}$ of the FS. This fact once
again demonstrates the pursuit of the FSs. In general, the crystal-like
structure \ can be seen more clearly if $\rho _{3}$ is observed as shown in
Fig.5c.

\qquad When N$_{out}$ is larger than 10, due to the rapid increase of $U$,
the associated (N$_{in})-$N$_{out}$ structure is no more dominant, and
therefore the crystal-like picture becomes ambiguous again due to the mixing
of structures. This is shown in Fig.4h and 4i.

\qquad \vspace{1pt}

6, ELECTRONIC STRUCTURES $\left( 91\leq L\leq 101\right) $

\qquad Inserting $j^{\prime }=1$ into eq.(16) and using the same argument as
before , it is straight forward to prove that the CFS of the $L=\frac{N(N-1)%
}{2}+N=91$ states include all the hollow (N$_{in})_{h}-$N$_{out}$ structures
ranging from N$_{in}=0$ to 12 . Therefore the (91)$_{1}$ would be nodeless
if the core is hollow. On the other hand, if the core is inert, all the (N$%
_{in}$)$_{c}$-N$_{out}$ are not the CFS (except the (12)$_{c}$-1). Thus, the
(91)$_{1}$ is expected to be hollow . This suggestion is confirmed by
Fig.3g. Similar to the (78)$_{1}$, the (91)$_{1}$ is also mainly contributed
by a single BF \{1-13\} with the weight 82.0\% . Due to the lack of coherent
mixing, the (91)$_{1}$ is liquid-like as shown in Fig.4j.

\qquad For the first-states with 92$\leq L\leq 101$ , we have

\qquad (i) The core-ring structure is explicit as representatively shown by
the $\rho _{1}$ plotted in Fig.3h to 3j. However, the core may be excited
and the probability of an electron staying at the center is smaller (3i and
3j).

\qquad (ii) \ It is noted that a state with a large $L$ would pursue a
larger moment of inertia to reduce the rotation energy. Since the structures
with N$_{out}$$<$ 7 have a smaller moment of inertia, these structures are
never found in the first-states with $L$ $\geq 93$ . \ Specifically, the (92)%
$_{1}$ is found to have N$_{out}$=7 as shown in Fig.4k.

\qquad (iii) Each of the (92)$_{1}$ , (94)$_{1}$ , (96)$_{1}$ , and (98)$%
_{1} $ states has only one FS, this FS has an appropriate N$_{out}$ , and
has \ $\stackrel{\_}{g}/$G$_{opt}$ \ \ $\approx 1$. \ Therefore these FSs
are competitive and are expected to be dominant. This point is confirmed by
the associated $\rho _{2}$ (cf. Fig.4), where a crystal-like picture with
the N$_{out}-$side polygons is seen. Furthermore, the N$_{in}$ of the FS of
the above four states are 6, 5, 4, and 3 (cf. Table 2), while the $N_{in}$
are 6.26, 5.01, 3.89, and 2.91, respectively . These values are one-to-one
close to each other. Thus, the pursuit of the FSs is further confirmed.
Besides, the FS of the above four states have $l_{in}=$ $(l_{in})_{b}$,
namely 15, 10, 6, and 3 (cf. Table 2), respectively. The corresponding $%
l_{in}$ calculated from $\rho _{1}$ are 19.70, 12.51, 7.38, and 4.19,
respectively. The latter set are always one-to-one bigger than the former
set due to having a slight core-excitation.

\qquad (iv) When $L\geq 100$ , the excited core (i.e., $l_{in}>(l_{in})_{b} $
) begin to compete seriously with \ the inert core. For the $L=100$ states,
the competing FSs are the (4)$_{h}$-9 and (2)$_{c}$-11 as shown in Table 2.
The $\stackrel{\_}{g}/$G$_{opt}$ of the former (latter) is considerably
smaller (larger) than one. It is noted that, when $L$ is large, the outer
ring would shift a little outward to increase the moment of inertia to
reduce the rotation energy. Thus,\ a small increase of $\stackrel{\_}{g}$\
is of advantageous, while a decrease of $\stackrel{\_}{g}$ is not. In fact,
it is the (2)$_{c}$-11 wins in the competition and is pursued by the
first-state, while the (4)$_{h}$-9 is pursued by the second-state. This is
shown in Fig.3h and 3i, and in Fig.4o and 4p. For the $(101$)$_{1}$, the (3)$%
_{h}$-10 is the only FS, and is expected to be dominant as shown in Fig.3j
and 5d.

\qquad (v) For the (93)$_{1}$ , (95)$_{1}$ , and (97)$_{1}$ , the N$_{out}$
of their FS are smaller than 7 and therefore is not competitive. Although
the (99)$_{1}$ has N$_{out}$=8, however its $\stackrel{\_}{g}/$G$_{opt}$ \
is quite small. Thus these four states do not have a competitive FS, and
therefore do not have a clear-cut geometric structure to pursue. They are
liquid-like. Nonetheless, their $\rho _{1}$ are more or less similar to
Fig.3h, thus they still have clear core-ring structures.

\qquad (vi) All the first-states with 92$\leq L\leq 101$ rotate uniformly,
they have $\gamma \approx 1$ except the (99)$_{1}$ and (101)$_{1}$, The FSs
of these two states have a considerably smaller $\stackrel{\_}{g}/$G$_{opt}$
(cf. Table 2). Thus, due to eq.(14), if they rotate uniformly the ring would
be too close to the core . To avoid being too close, $\omega _{out}$ would
decrease a little. In this way, although the rotation energy may increase a
little, the potential energy may thereby considerably decrease. This
suggestion is confirmed by the fact that their $\gamma $ is really smaller.
Incidentally, since the angular momentum $l_{out}$ is strongly constrained
by symmetry via eq.(9a), and therefore can not be adjusted freely , the
decrease of $\omega _{out}$ would cause an increase of $I_{out}$ via the
relation $l_{out}=I_{out}\omega _{out}$.

\vspace{1pt}

7, ELECTRONIC STRUCTURES ($L\approx 200)$

\ \ \ \ \ \ \ The main finding of the above study is the \vspace{1pt}pursuit
of the FSs . Does this experience work when $L$ is much larger? \ To clarify
this point we shall no more go to the states one-by-one, instead we choose
arbitrary a range 196$\leq L\leq 201$ for the studying. Let us first
evaluate the accuracy of the calculation in this range. \ E.g., the energies
of the (199)$_{1}$ state calculated with 6000, 9000, and 12000 BFs,
respectively, together with the $\alpha ,$ $N_{in},$ $l_{in}$ and $\gamma $
are listed in Table 3. \ One can see that, although the convergency is not
very good, it is qualitatively acceptable.\vspace{1pt}

Table 3 \ The energies and the quantities extracted from the $\rho _{1}$ of
the (199)$_{1}$ in accord with the increase of the number of BFs.

\begin{tabular}{|c|c|c|c|c|c|}
\hline
Number of BFs & $\alpha $ & $N_{in}$ & $l_{in}$ & $\gamma $ & $E((199)_{1})$
\\ \hline
6000 & 3.121 & 3.96 & 10.56 & 1.268 & 739.97 \\ \hline
9000 & 3.123 & 3.97 & 10.75 & 1.238 & 739.86 \\ \hline
12000 & 3.125 & 3.98 & 10.86 & 1.227 & 739.81 \\ \hline
\end{tabular}
\qquad

\qquad The FSs are shown in Table 4. \ The FSs\ with the core inert ($%
l_{1}=0 $) are found to have a too large $\overline{g}/$G$_{opt}$ , and
therefore \ are not listed. \ Whereas an excited core is pursued. \ On the
other hand, a highly excited core ($l_{1}\geq 6$) would lead to a too small $%
\overline{g}/$G$_{opt}$\ as shown in the table. Thus, too weak and too
strong core-excitation are both not appropriate.\vspace{1pt}

\vspace{1pt}\qquad Table 4 \ A continuation of Table 2 for the first- states
with $(196\leq L\leq $ 201).

\begin{tabular}{|l|l|l|l|l|l|l|l|l|l|}
\hline
& \multicolumn{5}{|l|}{Favorable structures and } & \multicolumn{4}{|l|}{
Quantities evaluated from} \\
& \multicolumn{5}{|l}{their related \ features} & \multicolumn{4}{|l|}{the $%
\rho _{1}$of the first-states} \\ \hline
L & FS & $l_{1}$ & $l_{in}$ & $\overline{g}$ & $\overline{g}/$G$_{opt}$ & $a$
& $N_{in}$ & $l_{in}$ & $\gamma $ \\ \hline
196 & (5)$_{h}$-8 & 6 & 40 & 1.5612 & 0.72 &  &  &  &  \\ \hline
196 & (3)$_{h}$-10 & 6 & 21 & 1.5811 & 0.59 & 3.16 & 3.75 & 13.88 & 1.11 \\
\hline
196 & (2)$_{h}$-11 & 4 & 9 & 1.9437 & 0.52 &  &  &  &  \\ \hline
197 & (5)$_{h}$-8 & 3 & 25 & 2.0736 & 0.96 & 3.48 & 4.80 & 24.61 & 1.05 \\
\hline
197 & (4)$_{h}$-9 & 5 & 26 & 1.7097 & 0.71 &  &  &  &  \\ \hline
197 & (3)$_{h}$-10 & 3 & 12 & 2.1506 & 0.76 &  &  &  &  \\ \hline
198 & (4)$_{h}$-9 & 3 & 18 & 2.1082 & 0.85 & 3.25 & 3.90 & 18.65 & 1.10 \\
\hline
198 & (2)$_{h}$-11 & 5 & 11 & 1.7581 & 0.47 &  &  &  &  \\ \hline
199 & (4)$_{h}$-9 & 1 & 10 & 2.8983 & 1.20 & 3.13 & 3.98 & 10.86 & 1.23 \\
\hline
199 & (5)$_{h}$-8 & 5 & 35 & 1.7113 & 0.80 &  &  &  &  \\ \hline
200 & (5)$_{h}$-8 & 2 & 20 & 2.3717 & 1.01 & 3.47 & 4.93 & 20.88 & 1.10 \\
\hline
200 & (3)$_{h}$-10 & 4 & 15 & 1.9235 & 0.70 &  &  &  &  \\ \hline
200 & (2)$_{h}$-11 & 6 & 13 & 1.6172 & 0.43 &  &  &  &  \\ \hline
201 & (3)$_{h}$-10 & 1 & 6 & 3.1225 & 1.10 & 3.00 & 3.06 & 7.27 & 1.24 \\
\hline
201 & (2)$_{h}$-11 & 1 & 3 & 3.4641 & 0.93 &  &  &  &  \\ \hline
201 & (4)$_{h}$-9 & 6 & 30 & 1.5916 & 0.66 &  &  &  &  \\ \hline
\end{tabular}
\vspace{1pt}

\qquad\ For the (196)$_{1}$ none of the FSs are superior (their $\stackrel{\_%
}{g}/$G$_{opt}$ are too small), \ therefore this state would have a strong
mixing of structures and would be liquid-like. Among the three FSs of the
(197)$_{1},$ the (5)$_{h}-8$ has its $\stackrel{\_}{g}$ \ closer to G$_{opt}$
, thus this FS is predicted to be dominant. \ Similarly, based on Table 4 ,
the \ (4)$_{h}-$9 \ is predicted to be dominant in (198)$_{1}$ and (199)$%
_{1} $ ,the \ (5)$_{h}-$8 \ is predicted to be dominant in (200)$_{1}$, and
the \ (3)$_{h}-$10 \ is predicted to be dominant in (201)$_{1}$. \ \ \ It
turns out that, for the case with a dominant FS, the predictions are nicely
confirmed by the calculation. E.g., the N$_{in}$ of the above FSs of the
(197)$_{1}$ to (201)$_{1}$ are 5, 4, 4, \ 5, and 3, while the corresponding $%
N_{in}$ extracted from $\rho _{1}$\ are 4.80, 3.90, 3.98, 4.93, and 3.06 . \
The $l_{in}$ of the above FSs are 25, 18, 10, \ 20, and 6, while the
corresponding $l_{in}$ extracted from $\rho _{1}$\ are 24.61, 18.65, 10.86,
20.88, and 7.27 . These values are amazingly one-to-one close to each other,
and thus the analysis based on the FSs is convincing. Furthermore, the
associated $\rho _{2}$\ and $\rho _{3}$\ confirm also the predictions. \
Representative examples are given in Fig. 4q, 4r, 5e, and 5f.\vspace{1pt}

\qquad\ It is noted that the (199)$_{1}$ and (201)$_{1}$ have a considerably
larger $\gamma $. \ On the other hand, their most competitive FSs have a
larger $\stackrel{\_}{g}/$G$_{opt}$. Thus, if these states rotate uniformly,
the ring would be too far away from the core (cf. eq.(15)). To avoid being
too far away, the ring rotates a little faster to reduce the moment of
inertia without altering $l_{out}$. This is the reason why they have a
considerably larger $\gamma $.

\qquad In general, when $L$ is large, the size of the system would increase,
the core-ring structures become more clear-cut. Besides, the core would have
a higher excitation. As a result, all these states are hollow as shown in
Fig. 3k, 3$l$, 5e and 5f.

\vspace{1pt}

\vspace{1pt}8, MAGIC NUMBERS

\qquad \qquad The above discussions demonstrate that, based on the FSs, the
structures of the first-states can be more or less predicted. In this
section we shall see that the energies are also strongly related to the FSs.
Let us go back to Fig.2 \ where platforms and shoulders are shown. A
platform starting at $L_{a}$ and ending at $L_{b}=L_{a}+1$ implies $%
E((L_{b})_{1})=$ $E((L_{a})_{1})+\hbar \omega _{o}$, i.e., the $(L_{b})_{1}$
is an c.m. excited state of the $(L_{a})_{1}$. This fact implies that the
internal energy ( the energy without the c.m. motion) of the $(L_{b})_{i}$
states are relatively higher. This is also the case if a shoulder appears.
In this case, $L_{a}$ is a candidate of a magic number (CMN). Evidently, if
the $(L_{a})_{1}$ has a competitive FS and the $(L_{b})_{1}$ does not have,
a CMN arises. For example, the (78)$_{1}$ is inherently nodeless and is able
to get access to all symmetric configurations , while the (79)$_{1}$ has
only one CFS (12)$_{c}$-1 which is unfavorable to binding. Thus the 78
appears as a CMN. Similarly, the (91)$_{1}$ is inherently nodeless (if the
core is excited), while the (92)$_{1}$ has only the (6)$_{c}$-7 ( which is
not competitive due to N$_{out}=7$), thus 91 is a CMN. The (111)$_{1}$ \ has
a competitive FS (3)$_{h}$-10 . Although the (112)$_{1}$ has two FSs, namely
the (6)$_{h}-$7 and (5)$_{h}$-8 , however the former has a small N$_{out}$
while the latter has a too small $\stackrel{\_}{g}/$G$_{opt}=0.78$ . They
are both not competitive, thus 111 is a CMN. Finally, The (118)$_{1}$ \ has
a number of competitive FSs, namely the (3)$_{c}$-10 , (5)$_{c}$-8 , and (4)$%
_{h}$-9 , while the (119)$_{1}$ has only one FS (6)$_{h}$-7 , which is not \
competitive due to having N$_{out}=7$. Thus 118 is a CMN. These examples
exhibit that the CMN\ can be more or less predicted.

\qquad

9, SUMMARY

\qquad The electronic structures of the first-states have been studied. By
an analysis of symmetry constraint and by performing numerical calculation,
we have obtained a clear picture of the core-ring structures. When $L$ is
small (78$\leq L\leq 83$), the core and ring are connected. \ When $L$ is
larger than 83, the core-ring structure becomes more and more explicit. \
When $L\leq 100$, the core remains inert (the (91)$_{1}$ is an exception).
When $L$ \ is larger, core excitation begins to compete. When $L$ \ is much
larger (say, $L\approx 200$), core excitation becomes dominant and the
states are hollow.

\qquad {\it The number of particles and the amount of angular momentum
contained in the core (ring) are not only determined by dynamics, but depend
seriously on symmetry constraint.\ Due to the constraint, for a given state,
it is advantageous to pursue a specific kind of structure, but
disadvantageous to pursue another kind.} This leads to Table 2 and 4, where
the favorable structures (FSs) of each state are listed.

\qquad The identification of the FSs is the main result of this paper. Based
on the FSs, the structures of the first-states can be predicted to a great
extent, the formation of crystal-like structure and the appearance of magic
numbers can be explained. In particular,{\it \ if a first-state has a
remarkably competitive FS (both the }N$_{out}${\it \ and }$\stackrel{\_}{g}/$%
G$_{opt}${\it \ are appropriate), the FS would be pursued, and the state
would be crystal-like and possess the associated (}N$_{in})-$N$_{out}${\it \
structure. If the }$L=L_{a}$ states{\it \ contain one or more than one
competitive FSs while the }$L=L_{a}+1$ states{\it \ do not contain, then }$%
L_{a}${\it \ is a CMN.}

\qquad The FSs \ can provide us an objective base for the further
classification of states. \ The states having the same FSs can be grouped
into a kind, e.g., all the $L=87$ , 96, 105, $\cdot \cdot \cdot $ contain a
single FS (4)$_{c}-$9, thus they belong to the same kind and their
first-states would have the same (4)$_{c}-$9 structure.

\qquad\ Although only a N=13 system is concerned in this paper, the idea,
the way of analysis, the qualitative results are quite common to the
2-dimensional systems with an attractive center. In fact, both this paper
and the previous ref.\lbrack 8\rbrack\ provide qualitatively similar
message. Thus, it is not doubted that the physical picture provided by these
two papers can be generalized to the systems with an even larger N. \ Where,
the identification of the FSs is again a key to understand the electronic
structures .

\qquad

\vspace{1pt}Acknowledgment: This paper is supported by the NSFC\ of China
under the grant No.90103028, No.10174098, and by a fund from the Ministry of
Education of China.

\vspace{1pt}

REFERENCES

1, L. Jacak , P. Hawrylak, A. W\'{o}js, {\it Quantum Dots }(Springer,
Berlin, 1998)

2, T. Chakraborty, {\it \ Quantum Dots} (Elsevier, Amsterdam, 1999){\it \ }

3, M.S. Kushwaha, \ Surface Science Reports, {\bf 41}, 1 (2001)

4, S.M Reimann and M. Manninen, \ {\it Rev. Mod. Phys}. {\bf 74}, 1283 (2002)

5, G.W. Bryant , {\it \ Phys. Rev. Lett}. {\bf 59}, 1140, ({\it 1987})

6, P.A. Maksym, H. Imamura, G.P. Mallon, and H. Aoki, {\it J. Phys.:
Condens. Matter}{\bf \ 12}, R299 (2000)

\vspace{1pt}7, \ U. Meirav, M.A. Kastner, and S.J. Wind, \ {\it Phys. Rev.
Lett}.{\bf \ 65}, 771 (1990)

8, \ S. Tarucha, D.G. Austing, T. Honda, R.J. van der Haage, and L.
Kouwenhoven, {\it \ Phys. Rev. Lett.} \ \ \ \ \ \ \ \ {\bf 77}, 3613 (1996)

9, H. Drexler, D. Leonard, W. Hansen, J.P. Kotthaus, and P.M. Petroff,{\em \
}{\it Phys. Rev. Lett}. {\bf 73}, 2252 (1994)

10, M. Fricke, A. Lorke, J.P. Kotthaus, G. Medeiros-Ribeiro, and P.M.
Petroff, {\it Europhys. Lett}. {\bf 36}, \qquad \qquad \qquad\ \ 197 (1996).

11, W.Y. Ruan, Y.Y. Liu, C.G. Bao and Z.Q. Zhang, 1995 {\it Phys. Rev}. {\bf %
B51} 7942\ (2000).

12, C.G. Bao, {\it \ Phys. Rev. Lett}. {\bf 79}, 3475 ({\it \ 1997}).

13, C.G. Bao, J. Phys. :Condens. Matter \ {\bf 14}, 8549 (2002)

14, C. de C. Chamon, and X.G. Wen, {\it Phys. Rev}. {\bf B49}, 8227, (1994)

15, H.M. Muller and S.E. Koonin, {\it Phys. Rev}. {\bf \ B54}, 14532, (1996)

16, E. Goldmann and S.R. Renn, {\it \ Phys. Rev.} {\bf B60}, 16611, ({\it %
1999 })

17, S.M. Reimann, M. Koskinen, M. Manninen and B.R. Mottelson, {\it Phys.
Rev. Lett}. {\bf 83}, 3270, (1999 )

18, C. Yannouleas and U. Landman, Phys. Rev. B {\bf 66}, 115315 (2002)

\vspace{1pt}19, C.G. Bao, \ Few-Body Systems, {\bf 13}, 41 (1992).

20, C.G. Bao and Y.X. Liu, Phys. Rev. Lett., {\bf \ 82}, 61 (1999)

21, C.G. Bao, W.F. Xie, and W.Y. Ruan, \ Few-Body Systems, {\bf 22}, 135
(1997)

22, T. Seki, Y. Kuramoto, and T. Nishino, J. Phys. Soc. Japan, {\bf 65},
3945 (1996)

23, Ruan W\ Y, Chan K S, Ho H\ P\ and Pun E\ Y\ B, 2000 \ {\it J. Phys.:
Condens. Matter} {\bf 12}, 3911

\vspace{1pt}

\vspace{1pt}

\vspace{1pt}Caption

Fig.1 $\rho _{1}(r{\bf )}$ of the first-state (82)$_{1}$with 6000 (a), 9000
(b), and 12000 (c) basis functions. The unit of length in this paper is $%
\sqrt{\hbar /m^{\ast }\omega _{0}}=194.71\stackrel{\circ }{A}.$

\ \ \ \

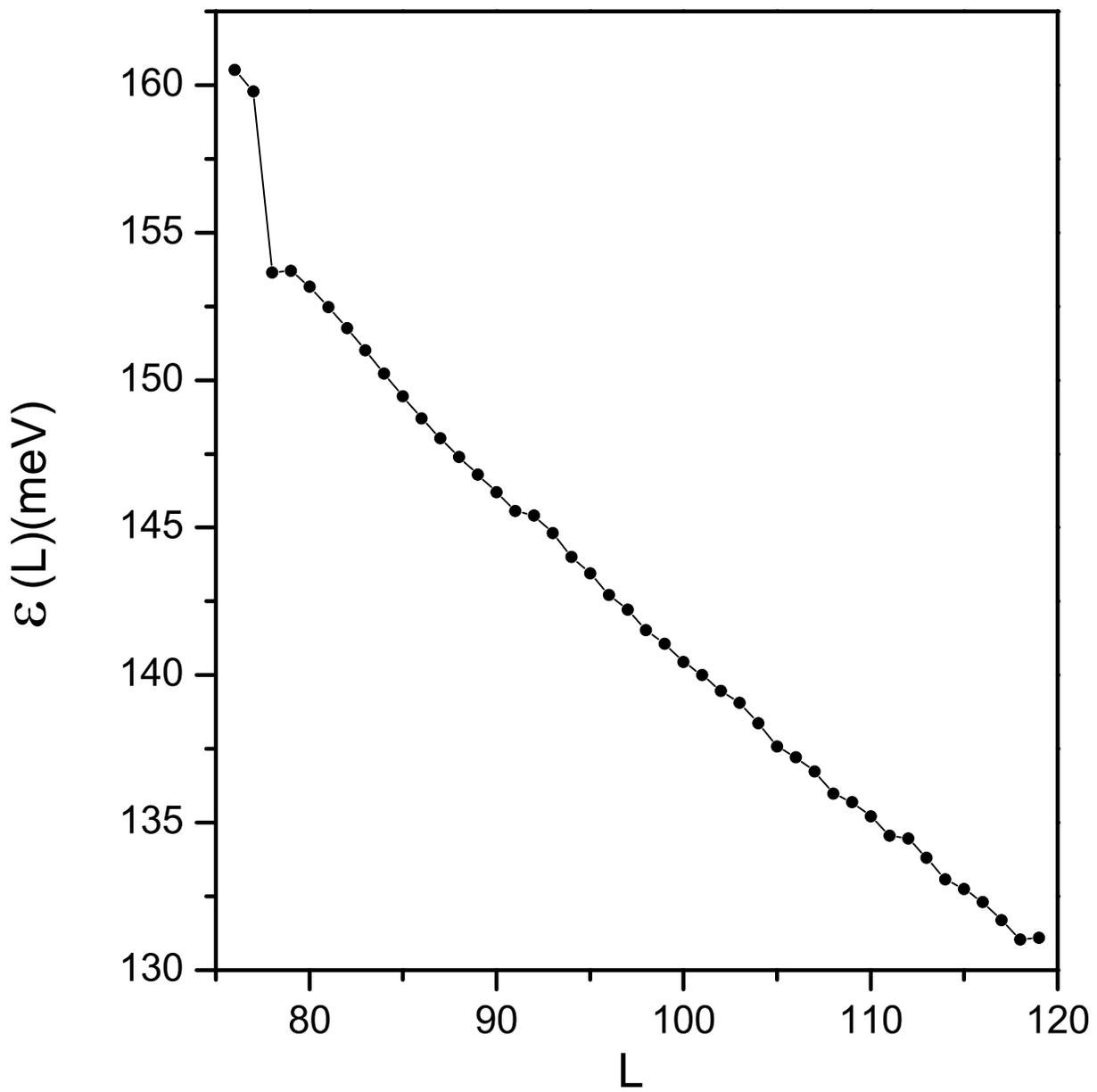
Fig.2 \ $\varepsilon (L)$ as a function of $L$. $\hbar \omega _{0}$=3meV is
assumed.

\vspace{1pt}

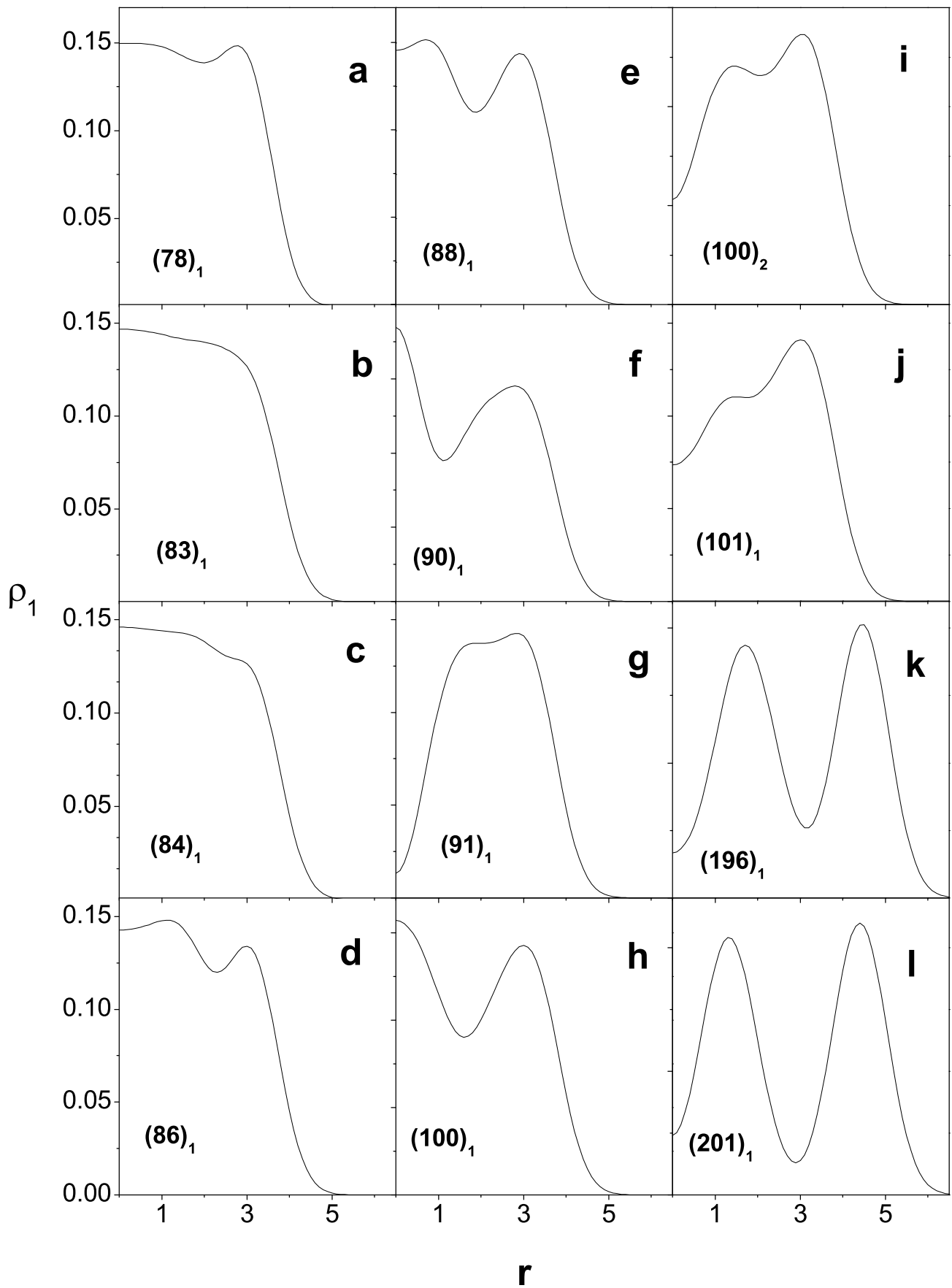
Fig.3 \ $\rho _{1}(r{\bf )}$ of the first-states (Fig.3i is for a
second-state).

\vspace{1pt}

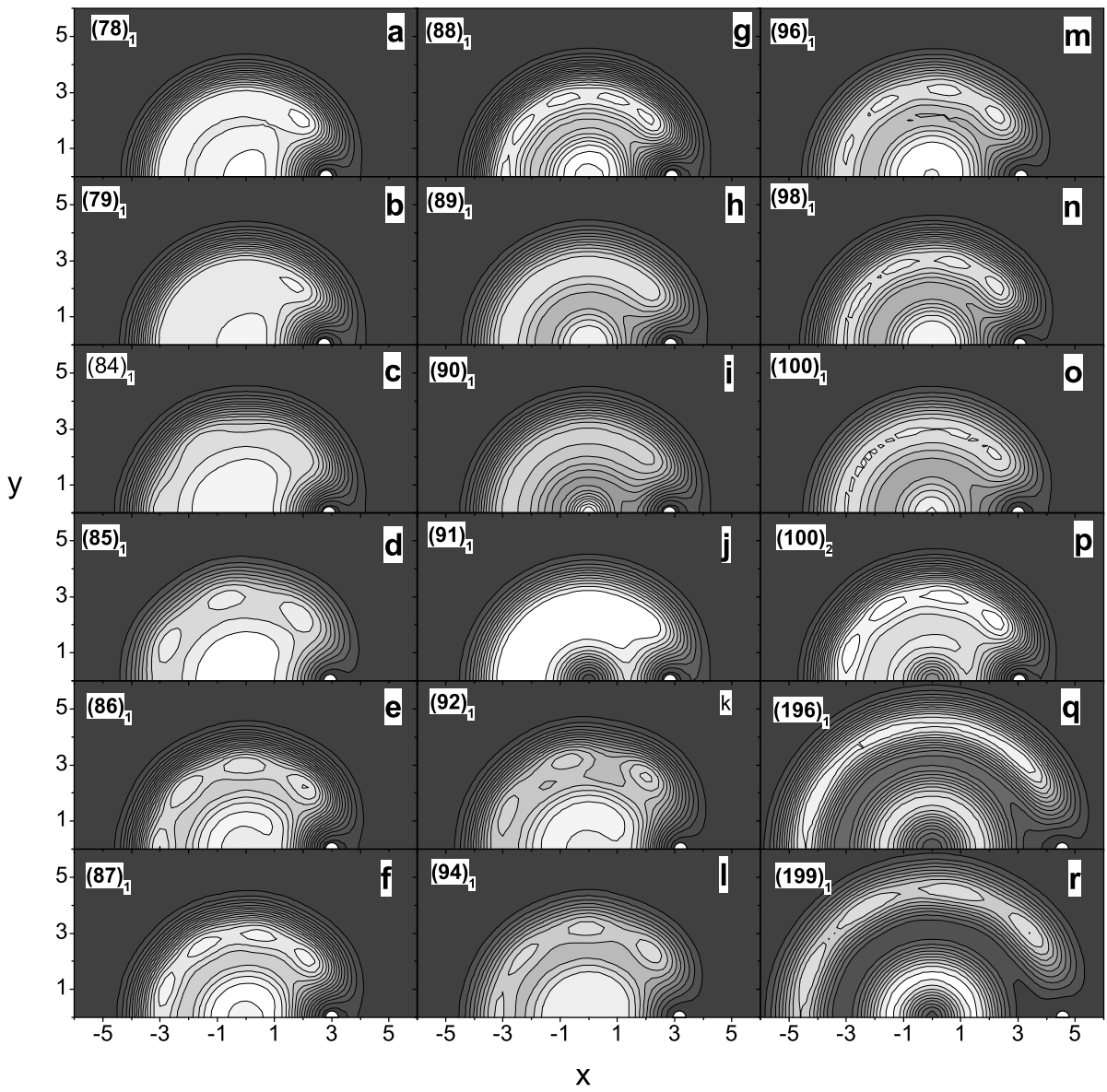
Fig.4 \ The contour plot of the two-body densities \ $\rho _{2}({\bf r,r}_{2}%
{\bf )}$ as a function of ${\bf r}$. The given ${\bf r}_{2}$ is marked by a
white spot. The lighter region has a larger \ $\rho _{2}$ .

\ \ \

\vspace{1pt}Fig.5 The contour plot of the three-body densities $\rho _{3}(%
{\bf r,r}_{2},{\bf r}_{3})$ as a function of ${\bf r}$. The given ${\bf r}%
_{2}$ and ${\bf r}_{3}$ are marked by two white spots. Refer to Fig.4.

\vspace{1pt}

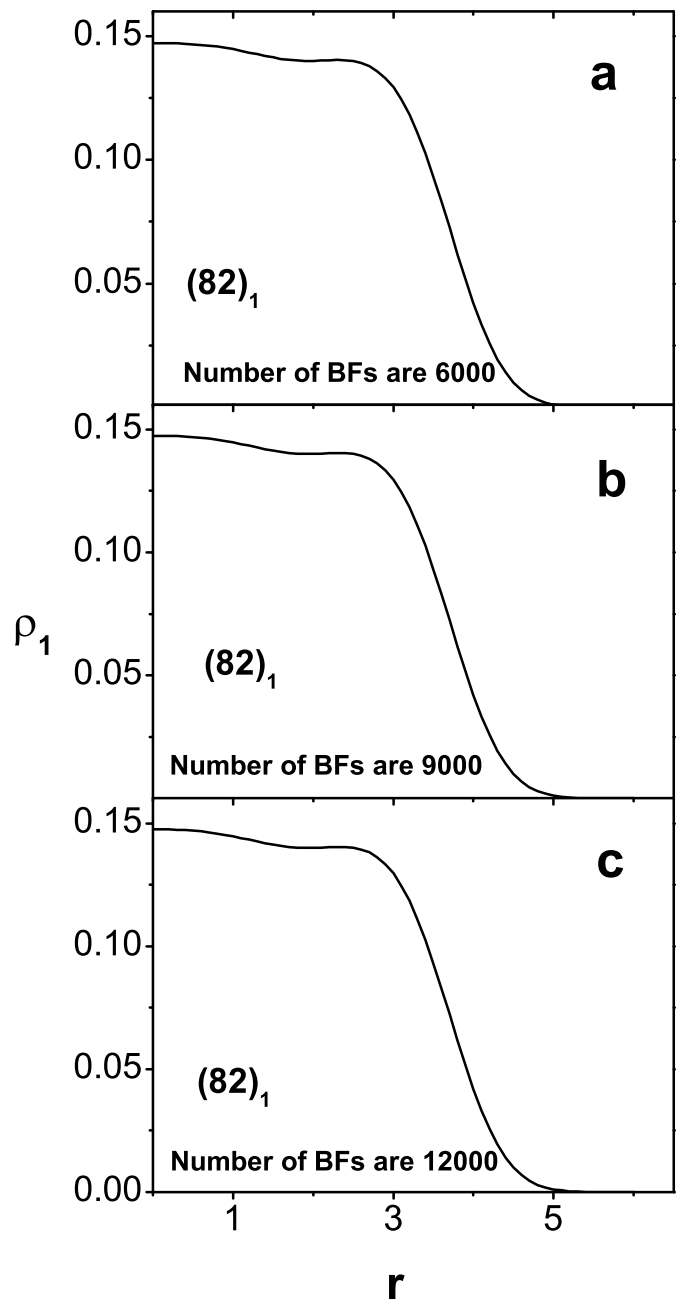
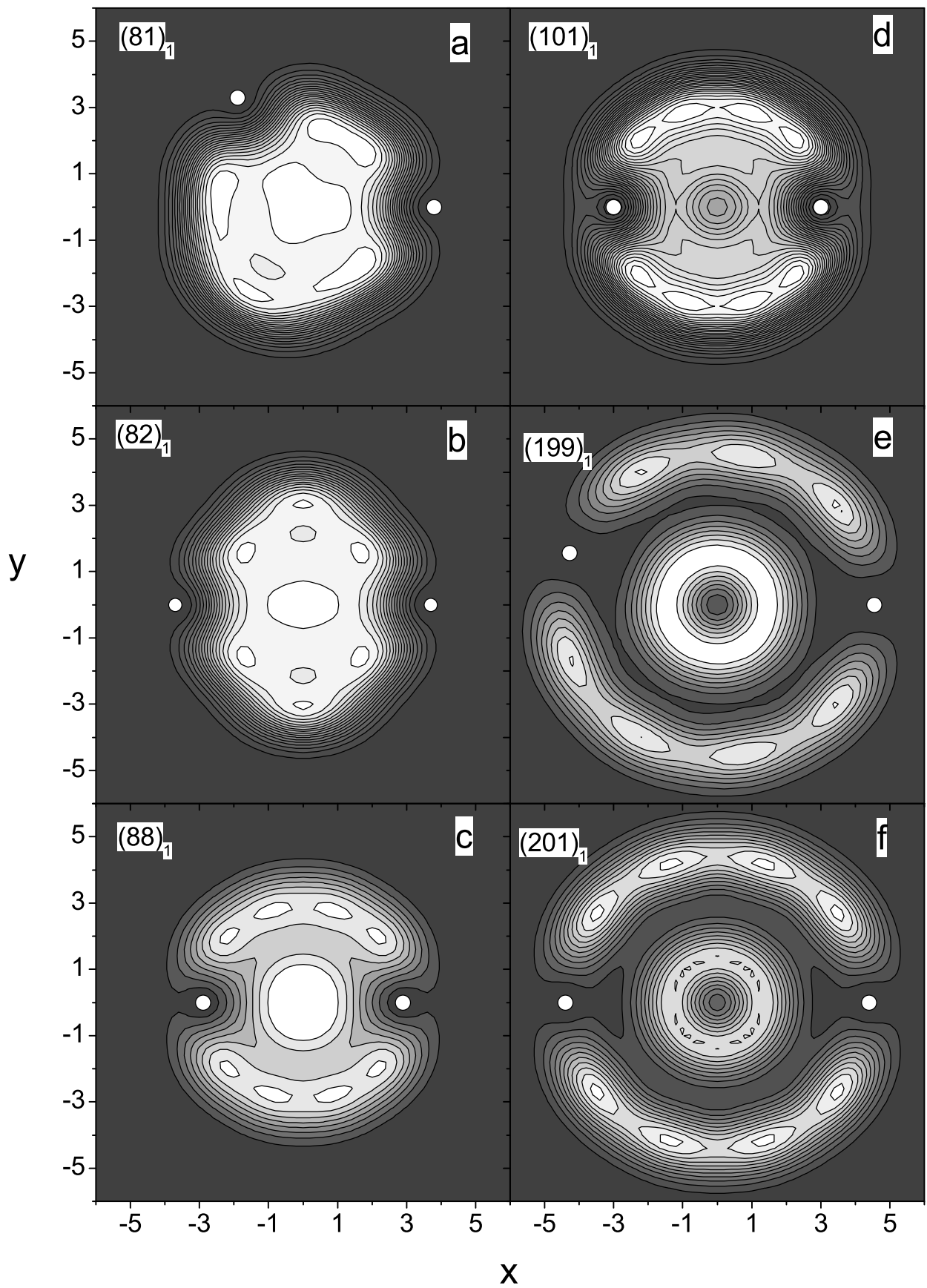
\end{document}